\documentclass[prb,twocolumn,showpacs,showkeys,preprintnumbers,amsmath,amssymb,superscriptaddress]{revtex4-1}

\usepackage{graphicx}
\usepackage{dcolumn}
\usepackage{bm}
\usepackage{color}

\begin{document}

\title{Continuity and boundary conditions in thermodynamics: \\From Carnot's efficiency to efficiencies at maximum power}

\author{Henni Ouerdane}\email{henni\_ouerdane@hotmail.com}
\affiliation{Russian Quantum Center, 100 Novaya Street, Skolkovo, Moscow region 143025, Russia}
\affiliation{Laboratoire Interdisciplinaire des Energies de Demain (LIED) UMR 8236 Universit\'e Paris Diderot CNRS 5 Rue Thomas Mann 75013 Paris France}
\author{Yann Apertet}
\affiliation{Institut d'Electronique Fondamentale, Universit\'e Paris-Sud, CNRS, UMR 8622, F-91405 Orsay}
\affiliation{Lyc\'ee Jacques Pr\'evert, 30 Route de Saint Paul, 27500 Pont-Audemer, France}
\author{Christophe Goupil}
\affiliation{Laboratoire Interdisciplinaire des Energies de Demain (LIED) UMR 8236 Universit\'e Paris Diderot CNRS 5 Rue Thomas Mann 75013 Paris France}
\author{Ph. Lecoeur}
\affiliation{Institut d'Electronique Fondamentale, Universit\'e Paris-Sud, CNRS, UMR 8622, F-91405 Orsay}

\date{\today}

\begin{abstract}
Classical equilibrium thermodynamics is a \emph{theory of principles}, which was built from empirical knowledge and debates on the nature and the use of heat as a means to produce motive power. By the beginning of the 20th century, the principles of thermodynamics were summarized into the so-called four laws, which were, as it turns out, definitive negative answers to the doomed quests for perpetual motion machines. As a matter of fact, one result of Sadi Carnot's work was precisely that the heat-to-work conversion process is fundamentally limited; as such, it is considered as a first version of the second law of thermodynamics. Although it was derived from Carnot's unrealistic model, the upper bound on the thermodynamic conversion efficiency, known as the Carnot efficiency, became a paradigm as the next target after the failure of the perpetual motion ideal. In the 1950's, Jacques Yvon published a conference paper containing the necessary ingredients for a new class of models, and even a formula, not so different from that of Carnot's efficiency, which later would become the new efficiency reference. Yvon's first analysis of a model of engine producing power, connected to heat source and sink through heat exchangers, went fairly unnoticed for twenty years, until Frank Curzon and Boye Ahlborn published their pedagogical paper about the effect of finite heat transfer on output power limitation and their derivation of the efficiency at maximum power, now known as the Curzon-Ahlborn (CA) efficiency. The notion of finite rate explicitly introduced time in thermodynamics, and its significance cannot be overlooked as shown by the wealth of works devoted to what is now known as finite-time thermodynamics since the end of the 1970's. The favorable comparison of the CA efficiency to actual values led many to consider it as a universal upper bound for real heat engines, but things are not so straightforward that a simple formula may account for a variety of situations. The object of the article is thus to cover some of the milestones of thermodynamics, and show through the illustrative case of thermoelectric generators, our model heat engine, that the shift from Carnot's efficiency to efficienc\emph{ies} at maximum power explains itself naturally as one considers continuity and boundary conditions carefully; indeed, as an adaptation of Friedrich Nietzche's quote, we may say that the thermodynamic demon is in the details.
\end{abstract}


\maketitle

\section{Introduction}
Thermodynamics is in our opinion a formidable and fascinating field of \emph{modern} physics. Some may find improper to relate the adjective ``modern'' to what is commonly perceived as a dusty 19th century phenomenological construct specifically designed for the making and maintenance of greasy and noisy steam engines, and superceded by the elegant and sophisticated statistical mechanics. As a matter of fact, this adjective is particularly relevant in at least two respects. First, from a historian viewpoint, the adjectives ``modern'' and ``contemporary'' relate to historical periods that cover the development of heat engines and the theories of heat. Next, and crucially, it is the undisputed validity of the laws of thermodynamics, which stand the test of time, that makes this field of thermal sciences still so timely. Albert Einstein once wrote about thermodynamics \cite{Einstein}:\\

``\emph{A theory is the more impressive, the greater the simplicity of its premises is, the more different kinds of things it relates, and the more extended is its area of applicability. Therefore the deep impression that classical thermodynamics made upon me. It is the only physical theory of universal content, which I am convinced that, within the framework of applicability of its basic concepts, will never be overthrown.}'' \\

\noindent Einstein's thoughts on thermodynamics and how this latter influenced some of his works, are discussed in Ref.\cite{Klein}. The view of Arthur Eddington on the matter is also interesting \cite{Eddington}:\\

``[\ldots] \emph{if your theory is found to be against the second law of thermodynamics I can give you no hope; there is nothing for it but to collapse in deepest humiliation.}''\\

\noindent This epigrammatic statement remains as strong as ever: thermodynamics, where applicable, has become a true touchstone for physical theories. Now, for all its strengths, thermodynamics remains a field of science. This means that not only its formulation has evolved, improved over time, and still does, but that its formalism and scope continually develops to account for a range of phenomena larger than those pertaining to equilibrium states. Indeed, inasmuch physical processes are characterized by irreversibility and nonequilibrium states, equilibrium thermodynamics may yield very incomplete information on the actual processes at work in a given situation. 

The building of nonequilibrium thermodynamics has also benefited from a phenomenological approach, which resulted in the force-flux formalism and Lars Onsager's recriprocal relations \cite{Onsager} in the early 1930s. One corner stone of the theory is the assumption of \emph{local} equilibrium. This formalism also rests on the hypothesis of minimal entropy production, which allows analyses in terms of quasi-static processes. These are important features of the theory since thermodynamic potentials may be well defined and the relationship between heat and entropy variations can be extended and assume a flux form. Nonequilibrium thermodynamics uses a terminology and provides tools to conveniently analyze transport and relaxation processes characteristic of steady-state out-of-equilibrium situations. To these processes are associated energy dissipation and entropy production, which is defined as the rate of variation of the total entropy given by the sum of the products of each flux by its conjugate affinity \cite{Pottier}. With the linear nonequilibrium theory one may perform optimization analyses but, as good and clever as it is, this theory does not offer such a practical or easily apprehensible framework for the study of actual heat engines and optimization of their operation. As a matter of fact heat engines had been essentially studied in the spirit of Sadi Carnot's approach \cite{Carnot}: improvement of the heat-to-work conversion, until the 1950s, as humanity was entering the so-called atomic age.

The discovery and control of nuclear fission at the end of the 1930s had been immediately put to use for military purposes in various countries; but given the tremendous costs this incurred, some governments also had to give way to the development of a civilian nuclear industry, to justify their spending of the public funds to the taxpayers who also happen to be voters. The tremendous power that could be wielded from atomic nuclei encouraged the countries engaged in such a technology to gather and discuss its future developments and applications to avoid disasters of unprecedented scale. The 1955 United Nations Conference on the Peaceful Uses of Atomic Energy offered the perfect forum for an appeased exchange of experimental results and theories on fission reactors, as well as views on the matter, thereby lifting \emph{to some extent} the shroud of secrecy imposed by the recent war times and by the then extreme political tension known as the cold war characterized by the sinister ``mutual assured destruction'' doctrine. It is in this context that one paper that contained a particular result that would later be considered as a paradigm shift in thermodynamics, was presented by Jacques Yvon, a French scientist. 

The name of Jacques Yvon is mostly associated to the BBGKY hierarchy in reason of his 1935 work on the statistical theory of fluids \cite{Yvonstat}. At the time of the 1955 Geneva Conference, Jacques Yvon was Head of the highly strategic D\'epartement d'Etudes des Piles (Department for the study of piles) at the Commissariat \`a l'Energie Atomique (Atomic Energy Commission) in Saclay, France. The work presented in his paper, ``\emph{The Saclay reactor: Two years of experience in the use of a compressed gas as a heat transfer agent}'' \cite{Yvon}, was concerned with the production of intense fluxes of neutrons in reactors and the handling of the accompanying large amounts of generated heat. As Yvon analyzed the use of compressed gases in reactors which could be designed to operate heat engines, he specifically mentioned the presence of a heat exchanger through which the reactors drive turbines. To compute the overall efficiency of the process, defined as the ratio of the power available on the turbine shaft to the heat flux transferred from the reactor's primary fluid to the working fluid, he separated the losses due to the system's imperfections from the thermodynamic efficiency of the working fluid. The temperature of the hot primary fluid, which serves as the hot thermal bath, was taken as a \emph{geometric average} of the maximum temperature that the system may withstand and the colder one of the condenser; such average, as Yvon stated, reflects the conditions yielding \emph{maximum power}. Hence, Yvon derived an expression almost as simple as that of the Carnot efficiency: the ratio of temperatures is replaced by the square root of this ratio, thus implying that also in this case the thermodynamic efficiency does not depend on the properties of the used working fluid. Interestingly, it does not depend on the heat exchangers' properties either.

With hindsight, we see that Yvon's analysis had some elements to engage in a reflection for a new framework for thermodynamic analysis. The object of the present article is thus to discuss and illustrate a genuine conceptual leap in the field of thermodynamics, which occured one century after Clausius published his statements of the first and second laws: the introduction of finite rates and hence time in the thermodynamic models, and the ensuing notion of \emph{objective}, which is here the maximization of output power. As reflected by the title of the article, accounting for time and rates amounts to addressing the fundamental questions of continuity and boundary conditions. Indeed, as in Yvon's study, the presence of heat exchangers between the thermal engine and the heat baths, as well as the search for conditions that maximize power, hence a quantity other than the conversion efficiency, have not for sole effect to more or less complicate the mathematical expression of the thermodynamic efficiency, but also and more importantly to modify the physical problem itself in a more profound fashion than one could expect at that time. Indeed, Yvon might have thought that his result was a mere by-product of a simplified optimization analysis, and he did not elaborate further on this. During the same period, Ivan Novikov, a soviet scientist, worked on similar problems and rederived Yvon's formula, crediting him for it, and stating that ``\emph{This formula is very useful for a rough calculation}'' \cite{Novikov}. But it took another fifteen years before Frank Curzon and Boye Ahlborn published a pedagogical paper \cite{ftt0}, which popularized so much Yvon's formula, that it is now called the Curzon-Ahlborn efficiency. By the end of the 1970s, other authors, including Bjarne Andresen, Peter Salamon and Stephen Berry, established the theory of finite-time thermodynamics, as a generalization of irreversible thermodynamics \cite{ftt1,ftt2,ftt3}.

The strong assumption made by Curzon and Ahlborn in their work was that the \emph{sole} source of irreversibility was through the coupling of the heat engine to its environment, while the working fluid was supposed to undergo a perfectly reversible transformation cycle. This particular configuration was named \emph{endoreversibility} by Morton Rubin who analyzed the optimal configuration for a class of heat engines undergoing finite cycle times \cite{Rubin}. The impression caused by the paper of Curzon and Ahlborn has been so strong that one could develop the misconception that finite-time thermodynamics reduces to the class of irreversible processes in the endoreversible configuation, and that the square root in the efficiency formula is its signature. In this article, considering endoreversibility as one limit, and exoreversibility \cite{Chen} as the other limit, we show that there is room for the intermediate situation as well \cite{PRE1}. The article is thus organized as follows. In Section 2, presenting some illustrative facts and dates, we try and show how some very practical problems inspired great minds in such a way that the ensuing flow of ideas, new concepts, and experimental data provided a very fertile space permitting the emergence of one of the most important fields of modern science. As the classical framework of thermodynamics was set, new questions, or rather new targets or objectives, arose. In Section 3, we review Carnot's model and we discuss the notion of efficiency constrained by the optimization of other quantities, through the critical examination of models leading to Yvon's formula. In Section 4, we see how the nature of irreversibilities influences the efficiency at maximum power, thus seeing that there is no universal upper bound to efficiency applicable to all cases of interest. We end the article with a discussion in Section 5.

\section{The 19th century: from heated debates to laws and order}

In the long history of engineering, a great variety of techniques and devices have been developed and designed to produce and control physical effects seen as desirable or \emph{useful}. Mechanical work is by far the most sought-after useful effect and heat engines occupy a particular place amongst the many human inventions. The fact that heat could be used to induce motion had been known for a long time but it was not until the 17th and 18th centuries that the technology based on steam, hence on heat, started to offer serious prospects for applications related to, e.g., the moving and transport of heavy objects. This partly owes to Boyle's law which offered a first proper understanding of the relationship between gas pressure and the volume occupied by the gas, and to the invention and improvement of thermometers. With the achievement of high-pressure steam engines thanks to the previous works of Thomas Newcomen, later followed by James Watt, this technology eventually imposed itself from the 19th century despite its main drawback: the production of mechanical work could require large quantities of fuel to obtain the necessary amount of heat, of which much was eventually \emph{lost}. This serious problem triggered an intense intellectual activity, which focused on the improvement and optimization of heat-to-work conversion devices, but more importantly gave birth to a very rich, fecund and far reaching field of science. 

While the notions of temperature and thermalization (now known as the zeroth law of thermodynamics) had been part of empirical knowledge for a long time, the concept of energy in its modern acception emerged only during the first half of the 19th century. This may seem astonishing considering the countless efforts, which had been devoted over the ages to the mastering of some of the forces of Nature, but even now the very essence of energy remains out of reach of our understanding; as Feynman put it in his lectures: ``\emph{It is important to realize that in physics today, we have no knowledge of what energy} is'', and this statement remains true. As a matter of fact, we know now that energy manifests itself and is stored under various forms, which were not all identified or even acknowledged as such at the turn of the 19th century, heat being the perfect example of a known natural phenomenon that was not properly understood. The prevailing view on heat amongst engineers and scholars at that time was that of Lavoisier's caloric theory: as a material substance which could neither be created nor destroyed, heat was a conserved quantity, spontaneously flowing from hotter to colder bodies.

The names of Thompson, Young, Mayer, Joule, Clausius and Helmholtz are often associated to the scientific and technical landmarks which permitted to establish $i$/ that heat and work are two modes of transfer of a quantity called energy, and $ii$/ that it is energy, but not heat, that is conserved. These two facts constitute the so-called first law of thermodynamics, one of the pillars of the then nascent thermal physics. As it turns out, however, another illustrious name must be associated to these scientific developments: that of Carnot. As he was reflecting on the \emph{motive power of heat}, Sadi Carnot based his studies on the caloric theory of heat and as he was greatly influenced by his father, Lazare, who was interested in finding ways of increasing the efficiency of water wheels, a  similarity between the flow water and the flow heat was evident to young Carnot. With hindsight, as Sadi Carnot's followed in his father's footsteps, one of the main merits of his approach was to have designed a \emph{generic model} of a heat engine, the Carnot engine, operating between two heat baths prepared at fixed temperatures, following an idealized \emph{reversible cyclic} process. The assumption of reversibility was important to ensure that the caloric could be recovered after one cycle (as would water getting back to its initial location in a water wheel), thus satisfying the (flawed) principle of its conservation. 

The strength of Carnot's analysis of heat-to-work conversion owes much to his building of a model devoid of the unnecessary details of particular engines. Indeed Carnot was not so much interested in the very nature of heat (he simply, but mistakenly, assumed that the caloric theory was correct), nor was he interested in heat transfer in material bodies or thermalization processes; rather he wanted to study heat as a \emph{cause} of motion. Some of his main conclusions are $i$/ that the motive power of heat solely depends on the two heat baths' temperatures: their difference triggers the flow of caloric, \emph{not} its consumption, or, in other words, work is a manifestation of the transport of caloric;  $ii$/ that the motive power decreases is some fashion as the temperature difference decreases; and $iii$/ that assuming slow reversible processes during the cycle, his model engine is the most efficient of heat engines operating between the same two heat baths. This latter conclusion is very important too: Carnot considered the \emph{conduction} of heat through the walls of the working fluid's container as heat leaks, and hence he chose not to include it in his model to find the perfect conditions for the working fluid to operate. As a matter of fact Sadi Carnot anticipated that (despite their equivalence in nature \emph{yet to be fully established}) heat and work could not strictly be treated on equal footing in the sense that a given amount of heat could never be all turned into work, while the opposite could, as shown by Thompson before \cite{Thompson}. 

Now, it is worthwhile to mention that in the same period in the 1820s, thermoelectricity, a then newly discovered physical effect related to heat, attracted much attention. Thermoelectric effects may be viewed as the result of the coupling between electrical transport and heat transport as these take place under suitable conditions. The names of Seebeck and Peltier are associated respectively to the rise of an electromotive force caused by a temperature difference across a conductor, and to thermal energy transfer, cooling or heating (but not heat production) at the junction of two conductors in which an electrical current circulates. These two thermoelectric effects are perfectly reversible in the sense that they are not associated to entropy production: only the Joule effect due to the electrical current is the source of entropy. In the 1840s, Joule published his results on the \emph{production} of heat in an isothermal conductor submitted to a voltage; this clearly was in fundamental disagreement with the then widely admitted property of conservation of caloric. As reported in Ref. \cite{Shaviv}, Joule's contradictors then used the reversibility of the Seebeck and Peltier effects as an argument to cast doubts on his conclusions, arguing that phenomena related to heat in electrical conductors were rather in favor of the caloric theory of heat. 

By the middle of the 19th century, many constitutive elements of the thermal science in gestation were published and debated, but it took the sharp minds of Clausius and Thomson (future Lord Kelvin, not to be confused with the aforementioned Thompson, Count of Rumford) to put together the relevant pieces of the thermal puzzle and form a clearly formalized and coherent theoretical construct, which would allow a proper interpretation of the wealth of experimental results gathered thus far, and make progress. Clausius' contributions to thermal sciences are such that he may be considered as one of the founding fathers of thermodynamics. Basing his analyses on the results of Mayer, Joule, and Carnot, Clausius gave a formulation of these, which clarified the situation and resolved fundamental contradictions. Thomson on the other hand used Carnot's analysis to propose the definition of an absolute temperature. He also built the framework which explained fairly consistently the laws of thermoelectricity. The second half of the 19th century then saw the maturation of the theory of heat, and the emergence of the kinetic theory of heat through the works of August Kr\"onig, Rudolf Clausius, James Clerk Maxwell, and Ludwig Boltzmann who, with Josiah Willard Gibbs, took the theory to unprecedented new heights.

\section{General considerations on continuity and boundary conditions}

\subsection{Carnot's model}
\begin{figure}
\centering
{\scalebox {0.32}{\includegraphics*{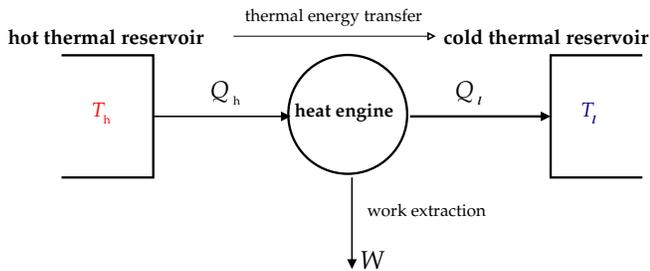}}}
\caption{Schematic representation of a heat engine operating between two heat baths at fixed temperatures $T_{\rm h}$ and $T_{\ell}$.}
\label{figCarnot}
\end{figure}

In this section we give a brief recap of the thermodynamics textbook classic: the Carnot cycle, the analysis of which differs much from the one proposed in the original work \cite{Carnot}. To picture the engine, consider an ideal gas at temperature $T_{\rm h}$ and contained in a chamber with the following characteristics: $i$/ the lateral sides are thermally isolated from the environment; $ii$/ one end side may be brought in thermal contact with a heat bath so that heat may be transferred (exclusively) to or from the gas, which acts as a working fluid; $iii$/ the other end side is closed by a mobile piston, which is also thermally isolated, and which undergoes no friction with the chamber walls as it moves. The operating heat engine may be schematically represented as on Fig.~\ref{figCarnot}, placed between two reservoirs, a hot one and a cold one, at fixed temperatures. 

In Carnot's model, the mechanical work is related to variations of the gas volume. If the gas expands, it pushes the piston which thus experiences a mechanical force: work is produced. As the gas is compressed by the piston: work is received by the gas. We now comment briefly on the succession of thermodynamic transformations, which Carnot found necessary to organize as \emph{cycles}. Work is necessarily associated to a change of state of the system; so if its production is organized as a process over a full cycle, this ensures that work is done only because of thermal effects and that the caloric is conserved since the system gets back to its initial state. It is also instructive to note that the assumption of reversibility is also necessary for Carnot's purposes: the system \emph{always} finds itself in thermodynamic equilibrium so that no dissipation may occur during the course of the cycle, thus ensuring that at the end of the cycle, the system will be exactly in the same state as it was at the start. 

The Carnot cycle consists of four \emph{reversible} steps: 1/ as the system is put in contact with a heat bath at temperature $T_{\rm h}$, the gas first undergoes an isothermal expansion; 2/ then the system is thermally isolated from its environment, and undergoes further expansion but adiabatically, which lowers its temperature to $T_{\ell}$ (with $T_{\ell}<T_{\rm h}$); 3/ next, the system is put in contact with the heat bath at temperature $T_{\ell}$ and it is isothermally compressed, thus releasing the heat $Q_{\ell}$ to this bath; 4/ finally, the system is compressed adiabatically, with the effect that it is back to its initial state at temperature $T_{\rm h}$. During the step 1, as heat is fed to the gas, it must expand to conserve its initial temperature. The isothermal expansion forces the piston to move back and make way, so pressure can only decrease. During the step 2, as the system is thermally isolated, the gas continues to expand, but as it cannot exchange heat, its temperature decreases down to the temperature of the colder heat bath; during the step 3, the gas is compressed by the piston and its pressure increases but as it is put in contact with the thermal bath, it can exchange heat and by doing so its temperature remains constant; during the step 4, the system is themally isolated again but as compression continues without exchange of heat, the temperature rises up to its initial value.

The standard presentation of Carnot's cycle is now based on the calculation of entropy variations, and hence it follows the spirit of Clausius' analytic treatment of the problem \cite{Clausius}. Besides the study of heat as a cause of motion, Carnot's model has been useful to understand that heat may not be completely converted into work. This is where the notion of thermodynamic efficiency comes in: how much work can be extracted from a given quantity of thermal energy in transit through the engine? The answer takes a rather simple form in the context of Carnot's model: the input heat is $Q_{\rm h}$ and the output work is $W = Q_{\rm h} - Q_{\ell}$, so the efficiency $\eta_{\rm C}$ may be defined as:

\begin{equation}\label{etaCQ}
\eta_{\rm C} = \frac{W}{Q_{\rm h}} = 1 - \frac{Q_{\ell}}{Q_{\rm h}}
\end{equation}

\noindent where the subscript ${\rm C}$ stands for ``Carnot''. The expression (\ref{etaCQ}) may be truly useful if it can be rewritten in terms of known and controlable quantities; here, these are the temperatures of the heat baths. Both heat quantities exchanged during one cycle, $Q_{\ell}$ and $Q_{\rm h}$, depend on the heat baths' temperatures, $T_{\ell}$ and $T_{\rm h}$; but they also explicitly depend on the volume of the working fluid, which varies during the expansion and compression steps. Hence the calculation of the ratio $Q_{\ell}/Q_{\rm h}$ requires not only the knowledge of $Q_{\ell}$ and $Q_{\rm h}$, but also of the works involved in the adiabatic expansion and compression of the working fluid: as one has four variables to deal with, one needs four equalities. This latter necessity reflects how work and heat are intertwined. From the explicit calculation of the heat and work exchanged during one cycle, it can be shown that \cite{Blundell}:

\begin{equation}\label{Clausius}
\frac{Q_{\ell}}{Q_{\rm h}} = \frac{T_{\ell}}{T_{\rm h}}
\end{equation}

\noindent This result is known as Clausius-Carnot equality from which we get:

\begin{equation}\label{etaCT}
\eta_{\rm C} = 1 - \frac{T_{\ell}}{T_{\rm h}}
\end{equation}

\noindent which is a positive quantity always smaller than one. 

The efficiency $\eta_{\rm C}$ can (mathematically) reach the limit value of 1 if $T_{\ell}$ equals 0 while $T_{\rm h}$ is finite, or if $T_{\ell}$ is finite while $T_{\rm h}$ is infinite, but both situations are obviously impossible to achieve. For an infinite temperature, the very existence of the heat engine has no meaning, so this case presents no interest whatsoever; and a strictly zero temperature cannot be obtained and maintained by any means: as of yet, the lowest temperature achieved is circa 450 pK for a Bose-Einstein condensate \cite{Leanhardt}, and this requires so much energy that the \emph{overall} efficiency for such a setup is far too small to have any practical interest. More importantly, Eq.~(\ref{etaCT}), is the basis of the so-called Carnot theorem, which states that under the same working conditions, the Carnot engine is the most efficient of all heat engines operating between two heat baths. This sets $\eta_{\rm C}$ as an absolute upper bound since the slightest source of irreversibility would make any heat engine certainly less efficient than a Carnot engine; therefore, the efficiency of any reversible heat engine is the Carnot efficiency. 

\begin{figure}
\centering
{\scalebox {0.32}{\includegraphics*{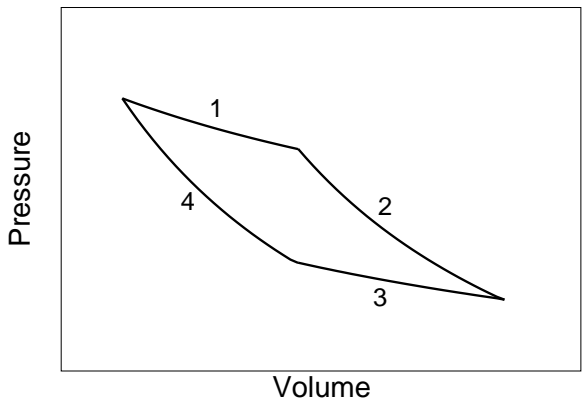}}}
\caption{Schematic representation of a pressure--volume diagram showing the four steps: isothermal expansion $(1)\longrightarrow$ adiabatic expansion $(2)\longrightarrow$ isothermal compression $(3)\longrightarrow$ adiabatic compression $(4)$ of a Carnot cycle.}
\label{figCarnotVP}
\end{figure}

Now, four remarks on Eq.~(\ref{Clausius}) are in order: $i$/ the ratio $Q_{\ell}/Q_{\rm h}$ does not depend on the volumes of the working fluid: this could be expected since at the end of the cycle everything has perfectly come back to its initial state, there is no memory of what happened, and the only pertinent quantities are the two heat bath temperatures; $ii$/ it can be rewritten as $Q_{\ell}/T_{\ell} = Q_{\rm h}/T_{\rm h}$, thus introducing the ratio heat/temperature known as \emph{entropy}, which is a state function. It has now appeared through the calculation of the thermodynamic efficiency that it is neither heat nor work that is conserved in Carnot's reversible cycle, but entropy; $iii$/ it is a particular case derived from a more general situation involving an inequality, which is known as the Clausius theorem \cite{Clausius}; $iv$/ Eq.~(\ref{Clausius}) involves temperatures, but we have not yet explicitly specified on which scale these are defined; as a matter of fact, the Clausius-Carnot equality has also been of great interest to define the \emph{absolute} temperature scale. As stated in Section 2, the notion of temperature had been known long before Carnot's work, and the first thermometer was invented more than two hundred years before Clausius' analyses; however, in the late 1840's, although he acknowledged that thermometers could yield extremely precise measurements, Thomson's view on thermometry was that its theory was unsatisfactory: ``[\ldots] \emph{we are left without any principle on which to found an absolute thermometric scale}'' \cite{Thomson}. Indeed, as all temperature scales made reference to specific materials (water, air, etc.) Thomson observed that: ``[\ldots] \emph{we can only regard, in strictness, the scale actually adopted as an arbitrary series of numbered points of reference sufficiently close for the requirements of practical thermometry}'' \cite{Thomson}. 

The solution to Thomson's quest for an absolute thermometric scale came from Carnot's model. It is of interest to note that Thomson shared at that time Carnot's view on the caloric: ``\emph{In actual engines for obtaining mechanical effect through the agency of heat, we must consequently look for the source of power, not in any absorption and conversion, but merely in a transmission of heat}'' \cite{Thomson}; but, more importantly, it is the fact that in Carnot's setup the mechanical work depends only on the temperature difference, and not on the properties of the working fluid, that put Thomson on the right tracks: he proposed a temperature scale on which each degree would correspond to the \emph{same} mechanical effect produced by one ``unit'' of heat in transit from the hot reservoir (at temperature $T_{\rm h}$) to the cold one (at temperature $T_{\ell} \neq T_{\rm h}$). In other words, for equal heat exchanges, one obtains equal temperature variations, irrespective of the properties of the working fluid, and one may thus write $Q_{\rm h} - Q_{\ell} = \Delta Q = \alpha \Delta T = \alpha T_{\rm h} - \alpha T_{\ell}$, with $Q_{\rm h} \neq Q_{\ell}$, which necessarily implies that heat is not conserved. Hence, with $Q_{\rm h} = \alpha T_{\rm h}$ and $Q_{\ell} = \alpha T_{\ell}$, one recovers Eq.~(\ref{Clausius}). Therefore, up to a multiplicative constant, the absolute temperature is uniquely defined and the ratio of temperatures $T_{\ell}/T_{\rm h}$, equal to $Q_{\ell}/Q_{\rm h}$, thus remains the same, as long as the temperatures are defined on the absolute scale. Thomson used an increment of 1 degree Celsius for an increment of 1 degree on the absolute scale, and found that what he called ``\emph{infinite cold}'' corresponds to a finite negative temperature on the scale of Celsius, around $-273^{\circ}$C \cite{Thomson}.

To conclude this section, we note that, notwithstanding its merits, Carnot's engine is absolutely unrealistic for essentially two reasons: $i$/ the thermal contacts between the working fluid and the heat baths are done abruptly over the cycle: Figure~\ref{figCarnot}, where the system is represented with its essential characteristics, shows no element with finite thermal resistance that may ensure the continuity of the potentials (temperatures) between the heat sources and the engine, and the presence of four cusps on the $PV$ diagram, Fig.~\ref{figCarnotVP}, demonstrate that passing from one step to the next is a singular process; $ii$/ it is impossible to arrange such a setup where there is neither a begining nor an end to the cycle! As a matter of fact, no one says how the system is prepared and put in contact with the heat sources, and at which stage of the cycle the whole process starts. One may argue that these questions do not matter much since, such as it is, the model already provides much insight into the physics of actual heat engines. This argument is not fully acceptable: Carnot's setup is off the arrow of time, and the thermodynamic science needs to be refined to better describe real engines operating under real working conditions: the operation of a heat engine in our world is only possible at the price of dissipative coupling of the system to the rest of the universe, and whatever useful thing is produced, the production rate matters much. 

\subsection{Maximizing power: the seed of a conceptual leap}
The principal objective of nuclear power plants is, as for the other kinds of thermal power plants, to produce electrical \emph{power}. One speaks about power rather than energy, as one is concerned with the rate at which electrical energy is transferred as it is produced. These notions relate to two things: the work produced per unit electric charge (potential) and the flow of these charges (electrical current); the product of these two quantities is the actual electrical power produced. Although electrical power is obtained by conversion of mechanical power at the electromechanical generator level, the thermodynamic process and its associated efficiency concern the flow of heated pressurized steam that acts on the turbine blades and induces the rotation of the turbine shaft. In his paper, Novikov noticed that \cite{Novikov}:\\

\emph{In contrast to conventional power stations the contribution of fuel costs to the cost of power produced in atomic stations is much less than other contributions, of which the greatest arises from the high capital cost. Therefore, an atomic power station will be most economic when its power output is at a maximum, since, in this condition the capital cost per kilowatt installed would be least.}\\

\noindent The economical constraint thus led to consider that the desirable target was not so much the traditional heat-to-work conversion efficiency alone, but rather the maximization of the ouput power, followed by the best achievable efficiency considering this new constraint. The short analysis proposed by Yvon \cite{Yvon}, rests on this point: he calculates the optimum thermodynamic efficiency assuming conditions that yield maximum power as a constraint. 

Now, it is possible to widen the scope of this analysis. Instead of focusing solely on the efficiency at maximum power, one may also reason in terms of objectives: increase of efficiency or increase of power. In the same spirit of the familiar saying: one cannot have a cake and eat it too, one cannot achieve both maximum power and maximum efficiency at the same time. Although this may seem frustrating at first glance, as one is brought to trade one quantity off for another quantity, this opens up a whole new field of play: finite-time thermodynamics, the corner stone of which ``\emph{is all about the price of haste and how to minimize it} as Andresen puts it in his recent review \cite{ftt4}. 

\subsection{Derivation and discussion of the Curzon-Ahlborn efficiency}
The discrepancy observed between the Carnot efficiency and the values obtained from actual heat engines such as, e.g., nuclear reactors, coal-fired and geothermal steam plants used to be explained as essentially due to heat leaks and friction of moving parts of the system \cite{clsthrm1,clsthrm2,clsthrm3}. Curzon and Ahlborn found instructive to turn their attention to another basic cause possibly lowering the heat engines' performance: the rate at which heat is transferred to and from the working fluid as the system is brought in thermal contact with the hot and cold heat baths \cite{ftt0}. They argued that, if one disregards other causes of dissipation, the isothermal steps during which the working fluid thermalizes with either of the heat baths have to take forever in order to achieve Carnot's efficiency, with a zero output power as an unavoidable consequence. They then argued that the cycle must be speeded up. But this should not be done carelessly: to maximize the rate of incoming heat from the hot bath to the working fluid, the temperature difference between both must be as large as possible; however, to maximize the rate of outgoing heat from the fluid to the cold bath, again the difference of temperatures must be as large as possible. To accomodate for both constraints, the working fluid temperature must be equal to the arithmetic average of both baths' temperature and remain constant. As a consequence, no work may be extracted from the working fluid, and one ends up with a situation of zero power with zero efficiency. The interest of the work of Curzon and Ahlborn is that they showed how one can find a situation between the two abovementioned extreme cases, where power could be nonzero and maximized, hence with a nonzero efficiency too.

\begin{figure}
\centering
{\scalebox {0.32}{\includegraphics*{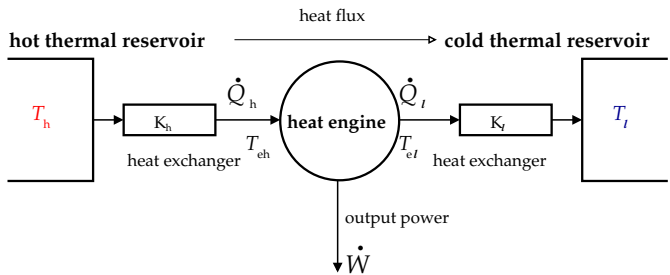}}}
\caption{Schematic representation of an endoreversible heat engine connected to two heat baths through heat exchangers with finite thermal conductance. The temperatures at the engine's ends are different from those of the baths.}
\label{figCAconf}
\end{figure}

The idea of Curzon and Ahlborn was to consider incoming and outgoing heat fluxes proportional to the respective differences of temperatures between the baths and those of the fluid across the container's walls. As a matter of fact, the walls play the r\^ole of effective heat exchangers with finite thermal resistance. What we call here a typical Curzon-Ahlborn configuration is depicted on Fig.~\ref{figCAconf}; the essential difference between this type of system and that of Carnot on Fig.~\ref{figCarnot} is the presence of heat echangers between the engine and what is represented as the cold and hot ends of the engines. This changes the nature of the problem: contrary to the Carnot setup, the resistances modify the boundary conditions in such a way that they ensure the continuity of the thermal potentials. As these exchangers are supposed to have a finite thermal conductance, the temperature difference experienced by the engine is different from the temperature difference between the hot and the cold heat baths. We denote $T_{\rm eh}$ and $T_{\rm e \ell}$, the hot and cold temperatures of the working fluid as it is put in thermal contact with the hot and cold baths respectively. Because of the presence of heat exchangers, we have $T_{\ell} < T_{\rm e \ell} < T_{\rm eh} < T_{\rm h}$. In the Curzon-Ahlborn model, the engine operates in a perfectly reversible fashion while energy is dissipated in the heat exchangers; as said earlier, this is called the endoreversible configuration \cite{Rubin}. 

With the assumption that the adiabatic steps of the cycle are reversible, Curzon and Ahlborn state that ``[\ldots] \emph{we must have}'':

\begin{equation}\label{ClausiusCA}
\frac{Q_{\rm e\ell}}{T_{\rm e\ell}} = \frac{Q_{\rm eh}}{T_{\rm eh}}
\end{equation}

\noindent which is an adaptation of the Clausius-Carnot relationship, Eq.~(\ref{Clausius}), to the case where the heat baths temperatures are replaced by the effective temperatures $T_{\rm e \ell}$ and $T_{\rm eh}$. The output power is given by $P = (Q_{\rm eh} - Q_{\ell})/t_{\rm tot}$, where $t_{\rm tot}$ is the duration of one full cycle of transformations. Differentiation of $P$ with respect to the temperature differences $T_{\rm e \ell}-T_{\ell}$ and $T_{\rm h}-T_{\rm eh}$ gives the conditions for maximum power, which involve the square root of the ratio of the heat baths' temperatures. The calculation of the thermodynamic efficiency under these conditions yields the following expression:

\begin{equation}\label{etaCA}
\eta_{\rm CA} = 1-\sqrt{\frac{T_{\ell}}{T_{\rm h}}}
\end{equation}

\noindent which has exactly the same form as Yvon's result \cite{Yvon}, is now known as the Curzon-Ahlborn efficiency. The comparison of their result to performance data of actual heat engines led Curzon and Ahlborn to conclude that: ``\emph{the result also has the interesting property that it serves as an accurate guide to the best performance of real heat engines}'' \cite{ftt0}. This latter statement contrasts with Novikov's view quoted in the Introduction: although the three authors agree on the usefulness of the formula, Novikov sees it as ``rough'' one. Equation (\ref{etaCA}) met with considerable success nonetheless in the thermodynamic community: it is simple, it bears some resemblance to Carnot's formula, and it shows no dependence on the properties of the working fluid nor on those of the heat exchangers. 

The square root of formula (\ref{etaCA}) could be seen as a signature of models that accounts for rates. However a number of studies also derive an efficiency formula with a square root while there is no rate in the models, and no mention of power either. For instance, in Ref.~\cite{maxW1}, the authors consider two heat baths and one heat source with \emph{finite} heat capacity $C$, which feeds the hot bath. There are three temperatures involved: $T^S>T^H>T^L$ (in their notations). The incoming heat is $C(T^S-T^H)$ and its conversion to work is assumed to be done with efficiency $(T^H-T^L)/T^H$; the produced work is the product of these latter quantities. Considering that $T^S$ and $T^L$ are fixed, but $T^H$ is variable, work is maximized for $T^H = \sqrt{T^ST^L}$, which is the geometric average of the two fixed temperatures. The authors thus found that the efficiency under the constraint of work maximization is that given by Yvon (they do not cite him though, and they do not cite Curzon and Ahlborn either). Although Yvon did not give much detail about his derivation in his paper \cite{Yvon}, we may safely assume that his reasoning probably was similar to that of the authors of Ref.~\cite{maxW1}, the nuclear reactor playing the r\^ole of the source here, and the primary fluid that of the hot thermal bath.

The efficiency at maximum work output was discussed in Ref.~\cite{maxW2} for the Otto, Joule-Brayton, Diesel, and Atkinson cycles, which deal with internal combustion heat engines; these systems differ much from the Carnot engine, but the author derived efficiencies equal or very close to $\eta_{\rm CA}$. His results came somewhat as a surprise to him since the studied models did not account for any finite-time process and for no irreversibility either. He came to conclude on this point that $\eta_{\rm CA}$ is a result more ``universal'' than previously thought, but that a reason for this could be that all the considered models are based on the strong assumption that over a complete cycle, the net entropy change of the working fluid is zero. In Ref.~\cite{maxW2}, the author also concludes on one conceptual difference between the standard Carnot model and those he presented; the former gives no clue as to what the operation of a heat engine entails: finding a compromise between objectives set by, e.g., economic, energy and time constraints, while the latter ones do, as illustrated by a work vs efficiency plot (we will get back to this consideration in the next section).  

The general nonequilibrium theory of reversible heat engines accounting for external irreversibilities (i.e., which occur outside the engines as these interact with their environment) is called ``\emph{endoreversible thermodynamics}''. By the end of the 1990's, various cycles, various heat transfer laws, various engines and system configurations had been treated in endoreversible thermodynamics \cite{endo}. And, concerning the universality of the Curzon-Ahlborn efficiency, it is interesting to note that the first derivation based on general arguments from linear irreversible thermodynamics was proposed in the year 2005 \cite{ftt5}; the motivation of the author of Ref.~\cite{ftt5} was to demonstrate that the formula for $\eta_{\rm CA}$, Eq.~(\ref{etaCA}), could be derived free of the assumption of endoreversibility, which he did for three setups: the standard one, a cascade with a continuum of auxiliary heat baths, and a tandem construction with one intermediate heat bath. The formula (\ref{etaCA}) then appeared to be a legitimate universal upper bound of thermodynamic conversion efficiency. But, in a paper published in 2008, a new formula different from that for $\eta_{\rm CA}$, was derived for a stochastic heat engine and also under quite general conditions \cite{exorv}. This triggered a reflection on the discrepancy, which will be discussed in the next section. 

Critical analyses are important and desirable to revise and refine the foundations of a theory, and discard what has come to appear plainly wrong. Two papers in particular have opposed some strong criticism to endoreversible thermodynamics. The first one by Dusan Sekulic in 1998, was entitled ``\emph{A fallacious argument in the finite-time thermodynamics concept of endoreversibility}'' \cite{ftt6}. The author disputes the validity of the Curzon-Ahlborn formula, arguing that the concept of endoreversibility, central to finite-time thermodynamics, is flawed. Indeed, Sekulic formally demonstrates that it makes no sense to consider that a heat engine is perfectly reversible while it is connected to components (heat exchangers) where irreversibilities take place. He also concludes that $\eta_{\rm CA}$ cannot be considered as an upper bound since it may take values lower than efficiencies of actual engines. In his Comment in the form of a rebuttal to Sekulic's paper, Andresen takes this opportunity to dissipate some confusion \cite{ftt7}. First and foremost, finite-time thermodynamics cannot be reduced to endoreversible thermodynamics, and does not rest on the hypothesis of endoreversibility. This latter was introduced to develop simple and illustrative cases for the theory. Next, he argues that a strict separation between the engine and its environment is mandatory to propose a clear picture of the phenomena at stake and hence to permit useful optimization studies. In his Reply \cite{ftt8}, Sekulic contends that his analysis holds and he reinforces his views with additional arguments proving that the rigorous definition of boundaries is key to a proper treatment of the thermodynamic problem, and that the concept of endoreversibility remains flawed as long as there is a coupling between a perfectly reversible engine and the dissipative elements of the considered thermodynamic system. Sekulic ends his paper with a challenge: ``\emph{So, the argument that there exists a logical fallacy in the concept of endoreversibility} [\ldots] \emph{still waits to be challenged}''.

The other paper that criticizes the approach of Curzon and Ahlborn presents the analysis of Bernard Lavenda \cite{ftt9}. The author reduces finite-time thermodynamics to the principle of maximum work as he concludes: ``\emph{The finite time thermodynamics that the Curzon-Ahlborn endoreversible engine supposedly introduces is illusory. The irreversibilities that supposedly arise from the coupling of the engine to the external world are fictitious. These statements are supported by the fact that Curzon and Ahlborn obtain the lowest final mean temperature achievable by which the system has done maximum work}''. Lavenda did not consider the discussion between Sekulic and Andresen, but it would have been profitable to refer to Andresen's Comment \cite{ftt8} to see that for its ``practitioners'', the general theory of finite-time thermodynamics does not rest on endoreversibility. However, Lavenda's analysis of the work of Curzon and Ahlborn is absolutely relevant and instructive in several respects. The first point is that Eq.~(\ref{ClausiusCA}) is wrong; as quoted above, Curzon and Ahlborn stated ``\emph{we must have} [\ldots]'', but this is a wrong educated guess, not an assumption that can be justified. For the assumption of reversibility to hold, the only valid equality is that of Clausius-Carnot, Eq.~(\ref{Clausius}). So, instead of basing the study on Eq.~(\ref{ClausiusCA}), Lavenda proposes the following: if $Q_{\rm h}$ is the heat absorbed by the engine, one may assume that the heat released is $Q_{\ell} = (T_{\rm m}/T_{\rm h})Q_{\rm h}$, where $T_{\rm m}$ is the mean working fluid temperature to be determined using Eq.~(\ref{Clausius}). The result, $T_{\rm m} = \sqrt{T_{\rm h}T_{\ell}}$, is the geometric average of $T_{\rm h}$ and $T_{\ell}$. Indeed, the two relationships between $Q_{\rm h}$ and $Q_{\ell}$ only yield to the basic definition of the geometric average \cite{defGav,means}. Sekulic's criticism of endoreversibility \cite{ftt6} thus finds an a echo in that of Lavenda who also addresses the key point of rates and hence the introduction of time in the model of Curzon and Ahlborn. Lavenda stresses that the introduction of two times, $t_{\rm h}$ and $t_{\ell}$ in our notations, should not represent a problem per se, but the problem is that it was further assumed that they are independent, while actually they are not: one goal of the study is to find conditions that maximize power, which is defined as the difference of the heat fluxes $Q_{\rm h}/t_{\rm h} - Q_{\ell}/t_{\ell}$, in a finite time, which remains indeterminate. Then Lavenda shows that the maximization problem had already been treated by Thomson himself, more than one century before in his paper ``\emph{On the restoration of mechanical energy from an unequally heated space}'' \cite{Thomson2}. Thomson considered a ``\emph{temperature to which the whole body can be brought by means of perfect engines, so that all the heat lost is converted into work}''; this temperature is the geometric mean of all the temperatures of the unequally heated body, and the time taken to achieve it being indeterminate, it cannot be used to yield a maximum power produced over a finite time. To end this section where we saw that the mathematical notion of means and their order has consequences on the building of models and the interpretation of their results, we mention a work by Edward Cashwell and Cornelius Everett \cite{meansTherm}, cited by Lavenda: in their study, the authors show mathematically how the first and second laws of thermdoynamics may be derived from the properties of means. Their basic model is that of an isolated system composed of a finite number of subsystems each with a mass, a heat capacity, and a temperature, initially placed in contact as the temperatures are not all equal, which is what Thomson called ``\emph{unequally heated space}''.

\section{Efficiencies at maximum output power}

\subsection{On the nature of irrversibilities}
\begin{figure}
\centering
{\scalebox {0.34}{\includegraphics{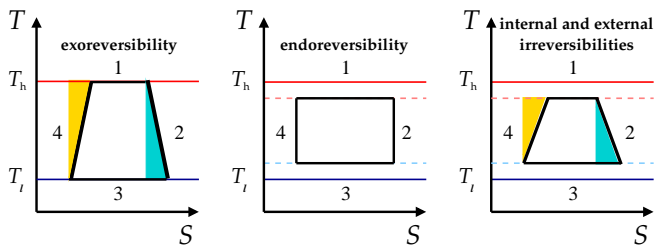}}}
\caption{TS diagrams under various situations of the nonideal Carnot cycle showing the four steps: isothermal expansion $(1)\longrightarrow$ adiabatic expansion $(2)\longrightarrow$ isothermal compression $(3)\longrightarrow$ adiabatic compression $(4)$. The heat generated by friction is represented by yellow triangular areas for the adiabatic step that raises the working fluid's temperature, and triangular blue areas for the adiabatic step lowering the fluid's temperature.}
\label{figTSdiag}
\end{figure}

In real heat engines, sources of irreversibilities are numerous; as examples, we may mention finite-rate (or, equivalently, resistance to) heat transfer, heat leaks, friction, combustion instabilities, etc. All sources of irreversibilities have an impact on engine performance, but their study shows that they usually have \emph{different} impact since they act in different fashions. As summarized in Ref.~\cite{Chen}, one may envisage essentially four situations to model thermodynamic cycles: 1/ external and internal reversibilities: Carnot cycle discussed in Section 3.1; 2/ external irreversibilities and internal reversibility: endoreversible configuration discussed in Section 3.3; 3/ external reversibility and internal irreversibilities: exoreversible configuration discussed below in Section 4; and 4/ external and internal irreversibilities: realistic configuration discussed below in Section 4. We analyze now the nature of irreversibilities and how they affect a system's performance, considering first an \emph{imperfect} heat engine undergoing a Carnot cycle. As discussed in Ref.~\cite{Wu}, the temperature-entropy $T$$-$$S$ diagram, fig.~\ref{figTSdiag}, where three different configurations for the nonideal Carnot engine are depicted, illustrates the fact that the adiabatic expansions are not always isentropic for a nonideal Carnot engine. 

The simplest situation is the endoreversible cycle. It looks very much like a Carnot cycle except that the presence of dissipative thermal contacts modifies the temperatures of the isothermal steps: these are not those of the heat baths, $T_{\rm h}$ and $T_{\ell}$, any longer but $T_{\rm eh}$ and $T_{\rm e\ell}$ (as defined in Section 3). In the exoreversible configuration, the temperatures during the isothermal steps are those of the heat baths, $T_{\rm h}$ and $T_{\ell}$. The imperfections manifest themselves during the adiabatic steps: As the piston moves during the adiabatic expansion or the adiabatic compression of the working fluid, friction against the chamber's walls dissipates energy: generation of heat accompanies the piston motion; but, as we assume that heat cannot cross these thermally insulated walls, it can only be transferred to the reservoirs during the isothermal steps of the cycle. This generated by friction heat is represented on Fig. \ref{figTSdiag} by triangular areas: yellow for the adiabatic step that raises the fluid's temperature from $T_{\ell}$ to $T_{\rm h}$ (step 4), and blue for the adiabatic step lowering the fluid's temperature from $T_{\rm h}$ to $T_{\ell}$ (step 2). It is important to realize that these two amounts of additional heat affect the engine performance quite differently: The heat produced during the step 4 reduces the amount of incoming heat flow during the isothermal step 1 at temperature $T_{\rm h}$, which yields an increase of the thermodynamic efficiency. Conversely, the heat produced during the adiabatic step 2 is rejected in the cold reservoir and cannot be recycled, which amounts to pure loss.

We now turn to the more realistic case with internal and external dissipations. It was already considered in Ref.~\cite{Wu}, where the authors introduced a quantity $R$, called the ``cycle irreversibility parameter'' and related to the ratio of two entropy differences, i.e. the ratio of useful heat production during step 4 to pure loss heat production during the step 2. Unfortunately, as it turns out, the definition of $R$ misses the modification of the temperatures at the engine's ends since the authors assumed that the temperatures of the isothermal steps remain the same as those for the endoreversible configuration. As a matter of fact, the heat produced during the adiabatic steps yields a slight but nonzero increase of both $T_{\rm eh}$ and $T_{\rm e\ell}$ when frictions are not neglected. The adiabatic steps thus are of great importance, but they are often neglected in the analysis of heat engines. Note that Novikov, in Ref. \cite{Novikov}, had already considered that internal dissipations may influence the performance of heat engines: he specifically introduced a parameter $A$ characterizing the distribution of the entropy produced by internal irreversibilities between the two adiabatic steps. However, in his paper, Novikov chose to focuse only an endoreversible model and thus neglected the effect of the parameter $A$.

\subsection{Thermoelectric generators as heat engine models of choice}
In a thermodynamic picture, thermoelectric generators are heat engines that use the conduction electrons as a working fluid. Hence, the conversion of a heat flux into electrical power (or vice-versa) in a device is a direct process since it is purely electronic in nature. As discussed in Ref. \cite{PRE3}, thermoelectric generators (TEG) are \emph{autonomous} heat engines: during their operation, the nonequilibrium steady-state is generated and maintained by time-independent boundary conditions imposed externally by, e.g., a load. Therefore, one may say that all the thermodynamic steps yielding output power, take place virtually at the same time in the various parts of the thermoelectric generators. But, interestingly, the equivalent thermodynamic cycle corresponding to the physical processes experienced by the electronic working fluid may be likened to a Carnot cycle: the exchanges of heat with the thermal reservoirs correspond to two isothermal steps while the charge transport across the thermoelectric generator and through the electrical load may be related to the two adiabatic steps. 

\begin{figure}
\centering
{\scalebox {0.175}{\includegraphics*{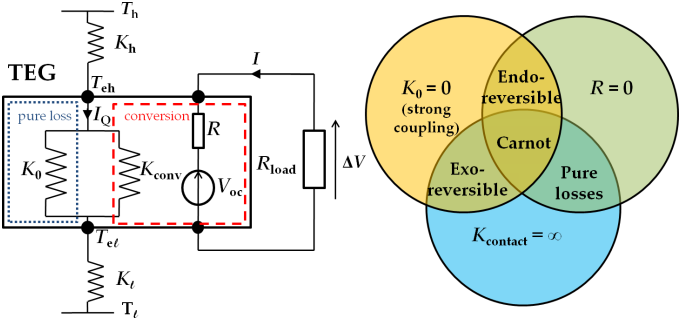}}}
\caption{Left: thermoelectric generator (TEG) dissipatively coupled to heat baths through heat exchangers with thermal conductance $K_{\ell}$ and $K_{\rm h}$ respectively. The thermoelectric effect permits the conversion of heat (heat flux $I_Q$) into electrical power $P=I\Delta V$ delivered to the load with electrical resistance $R_{\rm load}$. The TEG has an internal electrical resistance $R$, a thermal conductance resulting from the association in parallel of $K_0$ (heat leaks) and $K_{\rm conv}$ (Peltier term); the open circuit voltage is $V_{\rm oc}$. Left: diagram showing the various sources of irreversibilities and the  associated thermodynamic configurations.}
\label{figTEGirrev}
\end{figure}

In 1958 Sybren Ruurds de Groot was writing:``\emph{The phenomena of thermoelectricity have always served as touchstones for various theories of irreversible phenomena.}''  \cite{deGroot}. His statement has remained very true over the years as shown by the early 1990's work of Jeffrey Gordon \cite{Gordon} and some very recent ones covered in the next section. A simple model of thermoelectric generator (TEG) such as the one shown on the left side of Fig.~\ref{figTEGirrev}, has all the necessary ingredients to simply but rigorously describe the phenomena at stake. This figure shows a TEG dissipatively coupled to two heat baths at temperatures $T_{\rm h}$ and $T_{\ell}$ respectively, through heat exchangers with finite thermal conductances $K_{\rm h}$ and $K_{\ell}$. The effective temperatures on each side of the generator are $T_{\rm eh}$ and $T_{\rm e\ell}$. The equivalent thermal conductance of the exchangers is given by $K_{\rm contact}^{-1} = K_{\rm h}^{-1} + K_{\ell}^{-1}$. Heat transfer, as the flux $I_{\rm Q}$, within the generator may be modeled with two conductances in the parallel configuration: $K_0$ represents heat conduction which does not contribute to conversion and hence amounts to pure loss and is viewed as a parasitic process; $K_{\rm conv}$ is associated to conversion since it is the process that describes the transport of heat through the convective electrical current $I$ \cite{ApertetConv} which flows through the system and in the resistive load $R_{\rm load}$. The electrical part of the TEG is characterized by an open-circuit (Seebeck) voltage $V_{\rm oc}$ and its internal electrical resistance $R$.

\begin{figure}
\centering
{\scalebox {0.32}{\includegraphics*{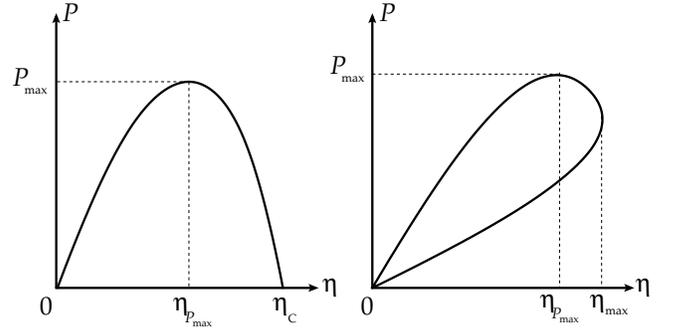}}}
\caption{Schematic representation of power versus efficiency curves for a generic heat engine without heat leaks (left) and one with heat leaks (right). The power is typically given by the work produced over a cycle time, and the efficiency is given by the ratio of the produced power to the heat input over the cycle time. These quantities may vary as the speed of operation of the heat engine is modified. The working conditions yielding maximum power do not correspond to those yielding maximum efficiency, $\eta_{\rm C}$ (without heat leaks) and $\eta_{\rm max}$ (with heat leaks). For a thermoelectric generator, the $P$ vs $\eta$ curve is parameterized by the electrical current $I$ delivered to the load; its intensity depends on the operating point, which is imposed by the load. The optimization strategy for the operation of a TEG depends on the desired target: power maximization or efficiency maximization, and also on the nature of the thermal constraints: constant temperature difference between the heat reservoirs or constant incoming flux in the TEG. These are analysed and discussed at length in Refs. \cite{ApertetEPL,ApertetJAP}.}
\label{figPvsEta}
\end{figure}

One may thus well identify three distinct sources of irreversibilities: the electrical resistance $R$ for the Joule heating (corresponding to mechanical friction), the effective thermal conductance $K_{\rm contact}$ of the heat exchangers between the heat baths and the engine, and the internal thermal conductance $K_0$ for heat leaks. Combination of all or of either of these sources yields to different models, which are summarized on the right side of Fig.~\ref{figTEGirrev}. In his paper \cite{Gordon}, Gordon studied the effect of these three types of irreversibilities on the relationship between the power output, $P$, and the conversion efficiency, $\eta$, of a thermoelectric generator. He showed that if Joule heating only is considered, the $P$ vs $\eta$ (parametric) curve resembles that of an endoreversible system (only dissipative thermal contacts) as schematically shown on the left side of Fig.~\ref{figPvsEta}. For a TEG, the curve is parameterized by the generated electrical current $I$ whose value depends on the operating point imposed by the load. Power generation occurs between the open circuit condition ($R_{\rm load}\equiv \infty$) and the short-circuit condition ($R_{\rm load} = 0$). In these two extreme cases the output power is zero, but in between there is an operating point where it reaches its maximum value. The open curve on Fig.~\ref{figPvsEta} may thus be described as follows: the point $(P=0, \eta=0)$ corresponds to the short-circuit while as one considers the limit $I\rightarrow 0$, one approaches the reversible limit associated to the Carnot efficiency $\eta_{\rm C}$ and vanishing power. Now, if one considers heat leaks, the $P$ vs $\eta$ curve qualitatively changes as it becomes a closed curve. This may be simply explained as follows: again, the point $(P=0, \eta=0)$ corresponds to the short-circuit; but as in the limit $I\rightarrow 0$ the power $P$ vanishes, so does the efficiency since heat leaks continue to dissipate energy irrespective of the vanishing current. Heat leaks thus act as a thermal by-pass, which implies that a finite amount of heat keeps being fed to the engine, even though it does not produce power: this is pure loss. Now, we saw that the sole internal dissipation (Joule heating here) yields an open $P$ vs $\eta$ curve characterized by an efficiency at maximum power $\eta_{P_{\rm max}}$, which is very much like the endoreversible (external dissipation) $P$ vs $\eta$ curve characterized by the Curzon-Ahlborn efficiency $\eta_{\rm CA}$; so one may wonder whether these two efficiencies are the same after all, and if a generic description of efficiency as maximum power is possible. 

\subsection{From endoreversibility to exoreversibility: a continuous path}
Heat leaks are not involved in the conversion process; so one may neglect these, with the assumption of strong coupling, and consider models with essentially two types of dissipation: internal and external. For a system such as the TEG depicted on Fig.~\ref{figTEGirrev}, its internal electrical resistance and the thermal resistance of the heat exchangers are well identified as the sources of irreversibilities and their study does not involve a complicated formalism. However, the comparison of their relative influence on a system's performance is not so straightforward since in principle both quantities do not characterize the same physical phenomena. But, as mentioned above, thermoelectric heat engines are the perfect systems for this purpose: the presence of heat exchangers manifests itself also in the definition of the effective electrical internal resistance of the operating TEG \cite{ApertetEPL,ApertetEPLR}, and this permits a physically transparent analysis of the phenomena at stake, their coupling, and their impact on the system's performance. 

Since its publication much emphasis was put on the Curzon-Ahlborn efficiency in finite-time and irreversible thermodynamic models of heat engine as shown by the literature cited in the present article. The 2008 paper of Tim Schmiedl and Udo Seifert \cite{exorv} came not only with a new formula for efficiency at maximum power, but also with a conceptual question: ``\emph{The Curzon-Ahlborn efficiency is a rather universal upper bound on the efficiency at maximum power for a wide class of heat engines. Why does the efficiency at maximum power derived here also under quite general conditions differ significantly from the Curzon-Ahlborn result?}'' These authors considered a particular version of a Carnot engine: a Brownian particle acting as a working fluid confined in a time-dependent trapping potential acting as a chamber, and derived the following expression for the efficiency at maximum power:

\begin{equation}
\eta_{\rm SS} = \frac{\eta_{\rm C}}{2 - \gamma\eta_{\rm C}}
\end{equation}

\noindent where $\gamma$ is a system-dependent parameter taking values between 0 and 1. Also in 2008, Zhan-Chun Tu derived an expression for the efficiency at maximum power of a Feynman ratchet \cite{ZCTu}, different from both $\eta_{\rm CA}$ and $\eta_{\rm SS}$. This latter article thus supported to some extent the idea that there is no such thing as a universal upper bound for the efficiency at maximum power but, as a matter of fact, an article which contained yet another formula in the form of an expansion:

\begin{equation}\label{etaemp}
\eta_{\rm PM} = \frac{\eta_{\rm C}}{2} + \frac{\eta_{\rm C}^2}{8} + {\mathcal O}(\eta_{\rm C}^3)
\end{equation}

\noindent was published in 2009 \cite{empexp}, indicating universality up to second order in the Carnot efficiency, for symmetric systems operating in the strong coupling regime. There was no physically transparent explanation for the discrepancy between $\eta_{\rm CA}$ and $\eta_{\rm SS}$ until very recently in 2012 \cite{PRE1}. 

The analysis developed in Ref.~\cite{PRE1} relied on a simple TEG model. The average temperature inside the TEG is $T'=(T_{\rm eh}+T_{\rm e \ell})/2$, with both $T_{\rm eh}$ and $T_{\rm e \ell}$ depending on the electrical current $I$ imposed by the external load \cite{PRE1,Yan}. As shown in Refs.~\cite{ApertetEPL,ApertetEPLR}, the effective internal electrical resistance comprises two contributions: one is the actual internal resistance $R$, while the other arises due to the presence of the heat exchangers and the thermoelectric coupling through the Seebeck coefficient $s$, and it is given by $R' = s^2T'/K_{\rm contact}$. As this latter effective resistance depends on the working conditions, it is convenient for the analysis to introduce a similar effective resistance, which does not: $R''=s^2T/K_{\rm contact}$ with $T=(T_{\rm h}+T_{\ell})/2$. The lower is $K_{\rm contact}$, the greater is $R''$, which is a measure of external irreversibilities.

\begin{figure}
	\centering
		{\scalebox {0.32}{\includegraphics*{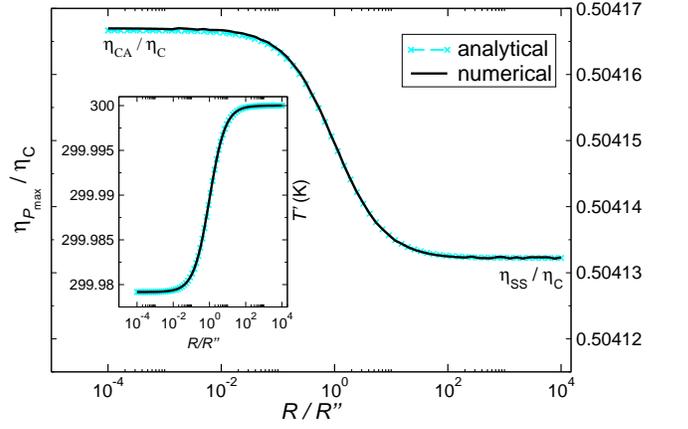}}}
	\caption{Efficiency at maximum power scaled to the Carnot efficiency and mean temperature $T'$ (inset) as functions of the ratio $R/R''$.}
	\label{fig:figure2}
\end{figure}

In the symmetric configuration the thermal contact conductances are equal: $K_{\rm h} = K_{\ell}$. In this case, the effective average temperature in the TEG operating at maximum output power is ~\cite{PRE1}:

\begin{equation}\label{tppmax}
T'_{P_{\rm max}}=T-\frac{R''}{R''+R}\frac{\Delta T^2}{16T},
\end{equation}

\noindent where $\Delta T=T_{\rm h}-T_{\ell}$. The second term on the right hand side of Eq.~\eqref{tppmax} characterizes the deviation from $T$ as the external dissipation becomes dominant, which in the limit $R''\gg 1$ permits to recover the Curzon-Ahlborn efficiency. The variation of $T'$ against the ratio $R/R''$ reflects the system's response when the internal dissipation constraint is relaxed. The insert of Fig.~\ref{fig:figure2} shows the excellent agreement between Eq.~(\ref{tppmax}) and the exact numerical calculation. 

Then, retaining only leading terms up to the third order in $\eta_{\rm C}$, the analytic expression for the efficiency at maximum power in the symmetric configuration reads \cite{PRE1}:

\begin{eqnarray}\label{eq:efficiencyPmax}
\eta_{P_{\rm max}}^{\rm sym} & = &\frac{\eta_{\rm C}}{2} \frac{1+\frac{\displaystyle \eta_{\rm C}}{\displaystyle 2(2-\eta_{\rm C})} \frac{\displaystyle R''}{\displaystyle R''+R}}{1-\frac{\displaystyle \eta_{\rm C}}{\displaystyle 4}\frac{\displaystyle R}{\displaystyle R''+R}}
\end{eqnarray}

\noindent For $R/R''\rightarrow \infty$, the sources of irreversibilities are essentially internal and  we get Schmiedl-Seifert efficiency $\eta_{P_{\rm max}}=\eta_{\rm SS}=\eta_{\rm C}/(2-\eta_{\rm C}/2)$, since $\gamma=1/2$ for this particular heat engine \cite{PRE1}; conversely, for $R/R''\rightarrow 0$, an expansion to the third order in $\eta_{\rm C}$ yields:

\begin{equation}
\eta_{P_{\rm max}}=\frac{\eta_{\rm C}}{2}+\frac{\eta_{\rm C}^2}{8}+\frac{\eta_{\rm C}^3}{16}+ {\mathcal O}(\eta_{\rm C}^4)
\end{equation}

\noindent which is the Curzon-Ahlborn efficiency expansion to the same order. In perfect agreement with Eq.~(\ref{etaemp}), which was derived in Ref.~\cite{empexp}, the coefficient at the second order in Carnot efficiency remains the same: 1/8, over the \emph{whole range} of variation of the ratio $R/R''$, and we see a continuous transition between the two limits. The efficiency at maximum power, $\eta_{P_{\rm max}}$, is plotted against the ratio $R/R''$ in Fig.~\ref{fig:figure2} for a TEG working between two heat baths at $T_{\ell} = $ 295 K and $T_{\rm h} = $ 305 K. The analytical expression fits very well to the numerical curve computed from exact formulas. The dissymetric case ($K_{\rm h} \neq K_{\ell}$) was also discussed in Ref.~\cite{PRE1}: the 1/8 coefficient also appears at the second order in $\eta_{\rm C}$ of $\eta_{P_{\rm max}}$ but higher order terms must not be neglected.

Therefore, contrary to what may be inferred from, e.g., Refs.~\cite{maxW2,ftt5,Gordon89}, the Curzon-Ahlborn efficiency is not a genuine universal upper bound on efficiency at maximum power of actual heat engines. The efficiency $\eta_{\rm CA}$ remains a fundamental result in the frame of linear irreversible thermodynamics \cite{ftt5}, but it applies only to theoretical endoreversible configurations. The efficiency $\eta_{\rm SS}$ is the $\eta_{\rm CA}$ counterpart for exoreversible configurations, for which dissipation is fully internal. The distinction between these two general forms of efficiency at maximum power pertaining two to different classes of model systems, thus lifts some uncertainties and confusion in finite-time thermodynamics analyses. It is also crucial to see that as the sources of irreversibilities are tuned from one limit to the other, the efficiency at maximum power of real heat engines varies \emph{continuously} between $\eta_{\rm SS}$ and $\eta_{\rm CA}$.

\section{Discussion and concluding remarks}
The word ``thermodynamics'' appeared as a reference to Thomson's dynamical theory of heat, where ``\emph{heat is not a substance, but a dynamical form of mechanical effect}''. However, as the theory is concerned with equilibrium states and transformations that do not account for time, the word ``thermostatics'' has also been used or even preferred, to make a clear distinction with nonequilibrium theories involving fluxes and time rates. But, even though time seems to play no r\^ole in classical equilibrium thermodynamics, it remains deeply rooted in its first and second laws: Conservation of energy follows from invariance with respect to translation in time (a consequence of Noether's first theorem \cite{Noether}), and irreversibility characterizes a ``preferred direction'', or asymmetry, of time. This latter is evidenced by the fact that the thermodynamic efficiency, a measure of the quantity of heat that cannot be made available for work, has an absolute upper bound, which is strictly smaller than one. 

The asymmetry of time may be illustrated by Eddington's concept of ``arrow of time'', which imposes a one-way direction for events \cite{Eddington}; in other words any system in the known Universe is subjected to the principle of causality: from its inception, to its evolution and its end. A non causal system, for which there would be absolutely no way to define a begining or an end, is outside of our Universe. The notion of isolated system in thermodynamics is thus a theoretical consideration only, useful for models but impossible to achieve, even at the quantum level where ``observation'' yields wave function collapse. As all systems connect in one way or another with their environment, they exchange ``something'' with it as they mutually act on each other, or interact. This interaction comes at the price of energy, which must be spent in the coupling. Therefore, there is no such thing as a perfect coupling, nor zero coupling. This latter fact led to the shift from the Newtonian paradigm to the quantum view of things: there is a postitive finite quantity called the quantum of action, $h$, which represents an absolute lower bound of all actions (or, equivalently, coupling) in the Universe. The key point of the quantum conceptual revolution is that interactions occur in a discontinuous fashion (one speaks of quantization; an insightful discussion is in Ref.~\cite{GCT}). The Newtonian limit $h\rightarrow 0$ is at the basis of useful models, still in use, but its close inspection shows that it completely misses some fundamental properties of the Universe. 

Finite interaction implies that absolute zero temperature (or ``\emph{infinite cold}'' as Thomson put it) is impossible, and that energy is always spent for the coupling of a system to its environment through resistive elements (in a broad sense). Dissipation thus comes at a necessary cost to ensure causality, but two remarks are in order here. The first one is that there is a fundamental equivalence between action and resistance. This was illustrated, e.g., by a recent analysis of Feynman's ratchet with thermoelectric transport theory \cite{PRE4}, which put forward two types of resistance to distinguish the dynamical behavior of the considered system from its ability to dissipate energy; these resistances have the dimension of an action. The next remark concerns the fact that some dissipation processes do not ensure causality \cite{PRE1,Gordon}: contrary to the internal and external sources or irreversibilities discussed in the Section 4.3, heat leaks on their own do not make a heat engine sensitive to the outside world, since the engine must be coupled to heat baths and has its working conditions imposed by an external load. Sekulic's criticism of endoreversibility is thus well founded and perfectly acceptable from a fundamental viewpoint \cite{ftt6} and it could also be extended to exoreversibility. But, there is no such thing as a perfect model of real systems, and limiting cases always provide useful insights as Carnot's model did, as long as these are discussed with their limits of validity. 

The steam technology, which emerged towards the end of the 18th century, triggered but also fed from the so-called industrial revolution, which had a massive and lasting impact on nearly all aspects of everyday life and its social organization. More than two hundred years later, as we face a number of tough economic, industrial, and environmental challenges with a looming energy crisis as a background, we still place some hopes in the heat-to-work conversion process, developing new tools for waste heat \cite{wstht} harvesting and transformation. Waste heat indeed represents a fantastic source of energy from which hundreds of TWh may in principle be captured per year. Various technologies, which find their origins in the 19th century, exist and are used today: heat exchangers for energy storage and routing, heat pumps, organic Rankine cycle, and thermoelectricity. Depending on the specific working conditions and the objectives, one technology may be viewed as more efficient than another; for instance, thermoelectricity is more appropriate in cases where the powers involved are below 1 kW. As a matter of fact, thermoelectric applications could be relevant over a range of electrical powers spanning ten orders of magnitude down to the $\mu$W. Their low conversion efficiency still prohibits the development of wide-scale applications \cite{Vining}, but great efforts are devoted to the breaking of the glass ceiling over performance \cite{Mori}, and original ideas may open new routes in this field \cite{Ouerdane}.

As humanity entered in the so-called atomic age, the proponents of the nuclear technology popularized the dream to wield the almost limitless energy that controlled self-sustained nuclear chain reactions can yield. A close look at what nuclear technology fundamentally entails, besides the problems related to radioactive wastes, leads to realize that one effectively deals with steam technology again, with nuclear fuel replacing coal. The paper presented by Yvon \cite{Yvon} at the Conference on the Peaceful Uses of Atomic Energy was remarkable for that it brought back to the center of a wider discussion on nuclear reactors, the very practical question of the efficiency at maximum output power of a steam engine. This question was studied again by Novikov \cite{Novikov}, but the efficiency formula with the square root, made its way thanks to the model of Curzon and Ahlborn, and retained their names instead of that of Yvon. As a matter of fact, Paul Chambadal, author of a thermodynamics book \cite{Chambadal} in 1949 is sometimes cited for the first derivation of the formula. This explains why sometimes the efficiency at maximum power is also referred to as such with a string of names: Chambadal, Yvon, Novikov, Curzon, and Ahlborn, attached to it. But it is of interest to note that it is the reading of Chambadal's book that very recently led some authors to publish a paper summarizing the ``history'' of the Curzon-Ahlborn formula \cite{fttn}, dedicated to Henri Reitlinger as he gets credited with the priority of the derivation of the formula in his 1929 book \cite{Reitlinger}. In his sharp criticism of the Curzon-Ahlborn model, Lavenda \cite{ftt9} leads the reader to realize that Thomson himself should get the credit for the square root formula. This makes two more names to add to the abovementioned ones.

The names of several eminent scientists and engineers have been mentioned in the present article, but it is obvious that we could not do justice here to the many people who participated in one way or another to the birth and development of thermal physics. That the mechanical equivalent of heat had been the object of a priority dispute between Joule and von Mayer, shows that recognition for priority seemed to be as important as the discovery itself. It still does, and perhaps more importantly than ever. As he was reflecting on the award of the Nobel prize to Glashow, Salam and Weinberg, John Clive Ward mentioned in his memoirs that the ``\emph{conflict between premature publication and the fear of being scooped was now endemic}'' \cite{Ward}. This statement also points out to the danger of rushing for publication to comply with a trend which favors a seemingly ever increasing rate for scientific progress. The fact that it took nearly 20 years between the results obtained by Yvon and Novikov, and their derivation in the extended model of Curzon and Ahlborn proposing the inclusion of transfer rates in the theory, plus another 25 to 30 years for the useful criticism of Sekulic and Lavenda, shows that maturation takes time.

We end the present article acknowledging the singularity of the 19th century in the history of the physical sciences \cite{IMN}. Building on centuries of empirical and tentative theories, using a heuristic approach and by rigorous application of Ockham's razor, great minds managed to put together in unified frameworks a body of knowledge that established the classical picture of the physical world. Electromagnetism and thermodynamics, perfect illustrations of the \emph{lex parsimoniae} principle, were truly epoch making theories, also providing the fertile ground for the scientific revolutions of the 20th century. As a matter of fact, any conceptual leap and the ensuing intellectual accomplishments rest on the sound knowledge and understanding of the current paradigm; one may thus say that there is no strict limit between periods of time when actual scientific progress is made: the ``scientific'' 19th century was born several years before 1801 and ended several years after 1900. There is a continuity of ideas and visions, which is ensured through communication between contemporaries and transmission to the next generations.

%

\begin{thebibliography}{}
\bibitem{defGav} Take two strictly positive numbers, say $a$ and $b$; there exists a strictly positive number $G$ that satisfies: $\frac{a}{G} = \frac{G}{b}$, which is the common ratio of a geometric sequence: $a$, $G$, and $b$; the number $G=\sqrt{ab}$, being the geometric average of $a$ and $b$. Note that the geometric average is smaller than the arithmetic average. See Ref.~\cite{means} for further detail on means and their properties.
\bibitem{wstht} Heat engines produce useful work, but they also reject heat in virtue of the second law of thermodynamics. As the rejected heat is a by-product with lower or no utility at all, it is called ``waste heat''.
\end{thebibliography}
%


\end{document}